\documentclass{ws-mpla}
\begin{document}
\markboth{J.-P. Chen, A. Deur and Z.-E. Meziani}
{Sum Rules and Moments of the Nucleon Spin Structure Functions}
%
%
\catchline{}{}{}{}{}
%
%
\title{Sum Rules and Moments of the Nucleon Spin Structure Functions}
\author{J.-P. Chen, A. Deur}
\address{Thomas Jefferson National Accelerator Facility\\
12000 Jefferson Avenue, Newport News, Virginia 23606, USA
\\
jpchen@jlab.org}
\author{Z.-E. Meziani}
\address{Department of Physics, Temple University\\
Philadelphia, Pennsylvania 19122, USA
}
\maketitle
\pub{Received ()}{}
\begin{abstract}
The nucleon has been used as a laboratory to investigate its own spin structure  and Quantum
Chromodynamics. New experimental  data on nucleon spin structure at low 
to intermediate momentum
transfers combined with existing high momentum transfer data offer a comprehensive picture
of the  transition region from the {\it confinement} regime of the theory to its   {\it
asymptotic freedom} regime. Insight for some aspects of the theory is gained by
exploring lower moments of spin structure functions and  their corresponding sum rules (i.e.
the Gerasimov-Drell-Hearn, Bjorken and Burkhardt-Cottingham). These moments are expressed in
terms of an operator-product expansion using quark and gluon degrees of
freedom at moderately  large momentum transfers. The sum rules are verified to 
good accuracy   assuming that no singular behavior of the structure functions is present at very
high  excitation energies.  The higher-twist contributions have been examined through
the moments evolution as the momentum transfer varies from higher to lower values.
Furthermore, QCD-inspired low-energy effective theories, which explicitly include chiral 
symmetry breaking, are tested at low momentum transfers.
The validity of these theories is further examined as  the momentum transfer  increases to
moderate values. It is found that chiral perturbation calculations agree reasonably well 
with the first moment of the spin structure function $g_1$   
at momentum transfer of 0.1 GeV$^2$ but fail to reproduce the neutron data in the case 
of the generalized polarizability $\delta_{LT}$. 
\end{abstract}
\section{Introduction}	
In the last twenty-five years the spin structure of the nucleon led to a very productive
experimental and theoretical activity with exciting results and new challenges\cite{Fji}. This
investigation has included a variety of aspects, such as
testing Quantum Chromodynamics (QCD), the theory of strong interactions, in its perturbative regime 
{\emph via} spin sum rules (like the Bjorken sum rule\cite{bjo66}) and understanding how the spin of the nucleon is
built from the intrinsic degrees of freedom of the theory, quarks and gluons. 
Recently, results from a new generation of experiments performed at
Jefferson Lab seeking to probe the theory in its non-perturbative 
and transition regimes
 have reached a mature state.  The low momentum-transfer results offer insight in 
a region known for the collective behavior of the nucleon constituents and their interactions.
In this region it has been more economical to describe the nucleon using effective degrees
of freedom like mesons and  constituent quarks rather than current quarks and gluons. 
Furthermore, 
distinct features seen in the nucleon response to the electromagnetic probe, depending on the 
resolution  of the probe, point clearly to different 
regimes of description, i.e. a scaling regime where quark-gluon correlations are suppressed 
versus a coherent regime where long-range interactions give rise to the 
static properties of the nucleon.

In this review we describe an investigation\cite{e94010-1,e94010-2,e94010-3,eg1p,eg1d,HTN,HTP,HTP2,BJ}
 of the spin structure of the nucleon through  
the measurement of the helicity-dependent photoabsorption cross sections or asymmetries using virtual photons  
across a wide resolution spectrum. These observables are used to extract the
spin structure 
functions $g_1$ and $g_2$ and evaluate their moments. These moments are powerful tools 
to test QCD sum rules and unravel some aspects of the quark-gluon structure of the nucleon. 

\section{Sum rules and Moments}

Sum rules involving the spin structure of the nucleon offer an important opportunity to study QCD. In recent years
the Bjorken sum rule at large $Q^2$ (4-momentum transfer squared) and 
the Gerasimov, Drell and Hearn (GDH) sum rule\cite{GDH,GDH2} at $Q^2=0$
have attracted large experimental\cite{ahr01,dutz03,dutz04} and theoretical\cite{dre01} efforts that have provided us with rich information. 
This first type of sum rules relates the moments of
the spin structure functions (or, equivalently, the spin-dependent
total photoabsorption cross sections) to the nucleon's static properties.
The second type of sum rules, such as the generalized GDH sum rule\cite{ji01,ans89} or
the polarizability sum rules\cite{dre03,dre04}, relate
the moments of the spin structure functions to real or virtual 
Compton amplitudes, which can be calculated theoretically. 
Both types of sum rules are based on ``unsubtracted'' dispersion relations
and the optical theorem\cite{BD}. 
The first type of sum rules uses one more general 
assumption,
such as a low-energy theorem\cite{low} for the 
GDH sum rule and Operator Production Expansion
(OPE)\cite{ope} for the Bjorken sum rule, to relate the Compton amplitude to a 
static property.

The formulation below follows closely Ref. \cite{dre03,dre04}. 
Consider the forward doubly-virtual Compton scattering (VVCS)
of a virtual photon with space-like four-momentum q, i.e., 
$q^2=\nu^2-{\vec q}^2 = -Q^2 < 0$, where $\nu$ is the energy
and $\vec q$ is the three momentum of the virtual photon.
The absorption of a virtual photon on
a nucleon is related to inclusive electron scattering. The inclusive
cross section, assuming parity conservation, contains four partial cross sections (or structure functions):
$\sigma_T$, $\sigma_L$, $\sigma_{TT}$, $\sigma_{LT}$, (or $F_1$, $F_2$, $g_1$,
$g_2$). The first two are spin-averaged, while the last two are spin-dependent.
In this review, we will concentrate on the spin-dependent ones.
In the following discussion, we will start with the
general situation, i.e., sum rules valid for all $Q^2$, then discuss the two
limiting cases at low $Q^2$ and at high $Q^2$.

Considering the spin-flip VVCS amplitude $g_{TT}$ and assuming it has an appropriate 
convergence behavior at high energy,  
an unsubtracted dispersion relation leads to the following equation for
$g_{TT}$:
\begin{equation}
{\rm Re}[{g}_{TT}(\nu,Q^2)-g^{pole}_{TT}(\nu,Q^2)]
=
(\frac{\nu}{2 \pi^2}){\cal P}\int^{\infty}_{\nu_0}\frac{K(\nu',Q^2)
\sigma_{TT}(\nu',Q^2)}{\nu'^2-\nu^2}d\nu', 
\end{equation}
where $g_{TT}^{pole}$ is the nucleon pole (elastic) contribution, ${\cal P}$ 
denotes the principal value integral and
$K$ is the virtual photon flux factor.
The lower limit of the integration $\nu_0$ is the pion-production threshold on
the nucleon.
A low-energy expansion gives:
\begin{equation}
{\rm Re}[g_{TT}(\nu,Q^2)-g^{pole}_{TT}(\nu,Q^2)]=
(\frac{2\alpha}{M^2})I_{TT}(Q^2)\nu+\gamma_{TT}(Q^2)\nu^3+O(\nu^5),
\end{equation}
with $\alpha$ the electromagnetic fine-structure constant and $M$ the 
nucleon mass. $I_{TT}(Q^2)$ is the coefficient of the $O(\nu)$ term of
the Compton amplitude.
Equation (2) defines the generalized forward spin 
polarizability $\gamma_{TT}(Q^2)$ (or $\gamma_0(Q^2)$ as it was used in Refs.
\cite{e94010-3,dre03}).
Combining Eqs. (1) and (2), the $O(\nu)$ term yields a sum rule 
for the generalized GDH integral\cite{dre01,ji01}:
\begin{eqnarray}
I_{TT}(Q^2) 
&=& {M^2 \over 4 \pi^2 \alpha} \int_{\nu_0}^{\infty}{K(\nu,Q^2) \over \nu}
{\sigma_{TT} 
\over \nu} d\nu \nonumber \\
 =&& \hspace{-0.5cm} {2M^2 \over Q^2} \int_0^{x_0} \Bigr [g_1(x,Q^2) - {4M^2 \over Q^2} 
x^2 g_2(x,Q^2)\Bigl ] dx
\label{eq:gdhsum_def1}
\end{eqnarray}
where $x = Q^2/2M\nu$ is the Bjorken scaling variable. 
As $Q^2 \rightarrow 0$, the low-energy theorem relates I(0) to the anomalous magnetic moment of the nucleon, $\kappa$, and 
Eq.~(\ref{eq:gdhsum_def1}) 
becomes the
original GDH sum rule\cite{GDH,GDH2}:
\begin{equation}
I(0) =\int_{\nu_0}^{\infty}{\sigma_{1/2}(\nu)-\sigma_{3/2}(\nu) \over \nu}
d\nu
 = -{2 \pi^2 \alpha \kappa^2 \over M^2},
\label{eq:gdh}
\end{equation}
where $\sigma_{1/2\,(3/2)}$ is the total
photoabsorption cross section with a projection of
$1 \over 2$ ($3 \over 2$) for the total spin along the direction of photon 
momentum, and $2 \sigma_{TT} \equiv \sigma_{1/2}-\sigma_{3/2}$.
The $O(\nu^3)$ term yields a sum 
rule for the generalized forward spin polarizability\cite{dre03,dre04}:
\begin{eqnarray}
\gamma_{TT}(Q^2)&=&
(\frac{1}{2\pi^2})\int^{\infty}_{\nu_0}\frac{K(\nu,Q^2)}{\nu}
\frac{\sigma_{TT}(\nu,Q^2)}{\nu^3}d\nu \nonumber \\
=&&\hspace{-0.5cm}\frac{16 \alpha M^2}{Q^6}\int^{x_0}_0 x^2 \Bigl [g_1(x,Q^2)-\frac{4M^2}{Q^2}
x^2g_2(x,Q^2)\Bigr ] dx. 
\end{eqnarray} 

Considering the longitudinal-transverse interference amplitude $g_{LT}$,
with the same assumptions, one obtains:
\begin{equation}
{\rm Re}[g_{LT}(\nu,Q^2)-g^{pole}_{LT}(\nu,Q^2)]=
(\frac{2\alpha}{M^2})Q I_{LT}(Q^2)+Q \delta_{LT}(Q^2)\nu^2+O(\nu^4) 
\end{equation}
where the $O(1)$ term leads to a sum rule for $I_{LT}(Q^2)$, which relates 
it to the $\sigma_{LT}$ integral over the excitation spectrum.
\begin{eqnarray}
I_{LT}(Q^2)&=&{M^2 \over 4\pi^2 \alpha}
\int^{\infty}_{\nu_0}\frac{K(\nu,Q^2)}{\nu}
\frac{\sigma_{LT}(\nu,Q^2)}{Q}d\nu \nonumber \\ 
= && \hspace{-0.5cm}{2M^2 \over Q^2}\int^{x_0}_0 x^2 \Bigl [g_1(x,Q^2)+g_2(x,Q^2)
\Bigr ] dx.
\end{eqnarray}
The $O(\nu^2)$ term leads to the generalized longitudinal-transverse
polarizability\cite{dre03,dre04}:
\begin{eqnarray}
\delta_{LT}(Q^2)&=&
(\frac{1}{2\pi^2})\int^{\infty}_{\nu_0}\frac{K(\nu,Q^2)}{\nu}
\frac{\sigma_{LT}(\nu,Q^2)}{Q \nu^2}d\nu \nonumber \\
=&&\hspace{-0.5cm}\frac{16 \alpha M^2}{Q^6}\int^{x_0}_0 x^2 \Bigl [g_1(x,Q^2)+g_2(x,Q^2)
\Bigr ] dx.   
\end{eqnarray}

Alternatively, we can consider the covariant spin-dependent VVCS amplitudes 
$S_1$ and $S_2$, which are related to the spin-flip amplitudes 
$g_{TT}$ and $g_{LT}$:
\begin{equation}
S_1(\nu,Q^2)=
{\nu M \over \nu^2+Q^2} \Bigr[g_{TT}(\nu,Q^2)+{Q \over \nu}g_{LT}(\nu,Q^2)
\Bigl ], \nonumber
\end{equation}
\begin{equation}
S_2(\nu,Q^2)=
-{M^2 \over \nu^2+Q^2} \Bigr[g_{TT}(\nu,Q^2)-{\nu \over Q}g_{LT}(\nu,Q^2)
\Bigl ].
\end{equation}
Dispersion relation with the same assumptions leads to
\begin{equation}
{\rm Re}[S_1(\nu,Q^2)-S^{pole}_1(\nu,Q^2)] =
{4\alpha \over M}I_1(Q^2)+ \gamma_{g_1}(Q^2)\nu^2
+O(\nu^4), 
\end{equation}
where the $O(1)$ term leads to a sum rule for $I_1(Q^2)$:
\begin{equation}
I_1(Q^2)= 
{2M^2 \over Q^2}\int_0^{x_0}g_1(x,Q^2)dx.
\end{equation}
The $O(\nu^2)$ term leads to the generalized $g_1$
polarizability:
\begin{eqnarray}
\gamma_{g_1}(Q^2)&=&
{16\pi \alpha M^2 \over Q^6}
\int^{x_0}_0 x^2 g_1(x,Q^2) dx   \nonumber \\
&=& \delta_{LT} + {2\alpha \over M^2 Q^2}\Bigr (I_{TT}(Q^2)-I_1(Q^2)\Bigl ).
\end{eqnarray}

For $S_2$, assuming a Regge behavior at $\nu \rightarrow \infty$ given by
$S_2 \rightarrow \nu^{\alpha_2}$ with $\alpha_2 < -1$, the unsubtracted
dispersion relations for $S_2$ and $\nu S_2$, without the elastic pole 
subtraction, lead to a ``super-convergence
relation'' that is valid for any value of $Q^2$,
\begin{equation}
\label{eq:bc}
\int_0^{1}g_2(x,Q^2)dx=0,
\end{equation}
which is the Burkhardt-Cottingham (BC) sum rule\cite{BC}. It
can also be written as
\begin{equation}
I_2(Q^2)={2M^2 \over Q^2}\int_0^{x_0}g_2(x,Q^2)dx={1 \over 4}F_P(Q^2)
\Bigl (F_D(Q^2)+F_P(Q^2)\Bigr ),
\label{eq:bc_noel}
\end{equation}
where $F_P$ and $F_D$ are the Pauli and Dirac form factors for elastic
e-N scattering.  

The low-energy expansion of the dispersion relation leads to
\begin{eqnarray}
&{\rm Re}&\bigl [\bigl (\nu S_2(\nu,Q^2)\bigr )-\bigl 
(\nu S^{pole}_2(\nu,Q^2)\bigr ) \bigr ] \nonumber \\
=&& \hspace {-0.5cm} 2\alpha I_2(Q^2) - {2\alpha \over Q^2}
\bigl (I_{TT}(Q^2)-I_1(Q^2)\bigr )\nu^2 
+{M^2 \over Q^2} \gamma_{g_2}(Q^2) \nu^4 + O(\nu^6),
\end{eqnarray}
where the $O(\nu^4)$ term gives the generalized  $g_2$ polarizability:
\begin{eqnarray}
\gamma_{g_2}(Q^2)&=& {16\pi \alpha M^2 \over Q^6}\int_0^{x_0}x^2 g_2(x,Q^2)dx
\nonumber \\
=&&\delta_{LT}(Q^2)-\gamma_{TT}(Q^2)+{2\alpha \over M^2 Q^2}
\Bigl (I_{TT}(Q^2)-I_1(Q^2)\Bigr ).
\end{eqnarray}

At high $Q^2$,
the OPE\cite{SV,stein,JU} for the VVCS amplitude leads to
the twist expansion:
\begin{equation}
\label{eq:Gamma1}
\Gamma_1(Q^2)
\equiv \int_0^1 g_1(x,Q^2)dx
=\sum_{\tau=2,4,...} {\mu_\tau(Q^2) \over (Q^2)^{(\tau-2)/2}}
\end{equation}
with the coefficients $\mu_\tau$ related to nucleon matrix elements
of operators of twist $\leq \tau$.
Here twist is defined as the mass dimension minus the spin of an
operator and $\mu_\tau$ itself is a pertubative 
series in $\alpha_s$, the effective strong coupling constant.
Note that the application of the OPE requires summation over all
hadronic final states, including the elastic at $x=1$.

The leading-twist (twist-2) component, $\mu_2$, is determined by
matrix elements of the axial vector operator
$\bar\psi \gamma_\mu \gamma_5 \psi$, summed over quark flavors, where
$\psi$ are the quark field operators.
It can be decomposed into flavor triplet ($g_A$), octet ($a_8$) and
singlet ($\Delta\Sigma$) axial charges,
\begin{eqnarray}
\label{eq:mu2}
\mu_2(Q^2)
&=& 
  \left( \pm {1 \over 12} g_A\ +\ {1 \over 36} a_8 \right)
+ {1 \over 9} \Delta\Sigma\ +O(\alpha_s(Q^2)) ,
\end{eqnarray}
where +(-) corresponds to proton (neutron) and
the $O(\alpha_s)$ terms are the $Q^2$ evolution due to the QCD radiative
effects that can be calculated from perturbative QCD.
The triplet axial charge is obtained from neutron
$\beta$-decay, 
while the octet axial
charge can be extracted from hyperon weak-decay matrix elements assuming
SU(3) flavor symmetry. 
Within the quark-parton model $\Delta\Sigma$ is 
the amount of nucleon spin carried by the quarks. Deep Inelastic Scattering 
(DIS) experiments 
at large $Q^2$ have extracted this quantity through a global analysis 
of the world data\cite{Fji}. 

Eqs. (\ref{eq:Gamma1}) and (\ref{eq:mu2}), at leading twist, lead to the Ellis-Jaffe sum rule\cite{EJ} with the assumptions of SU(3) flavor symmetry and an
unpolarized strange sea.
The difference between the proton and the neutron gives the flavor non-singlet
term:
\begin{equation} 
\Gamma_1^p(Q^2)-\Gamma_1^n(Q^2)={1 \over 6}g_A+O(\alpha_s)+O(1/Q^2),
\label{eq:genBj}
\end{equation}
which becomes the Bjorken sum rule at the $Q^2\rightarrow \infty$ 
limit. 

If the nucleon mass were zero, $\mu_\tau$ would contain only a twist-$\tau$ operator. The non-zero nucleon mass induces contributions to $\mu_\tau$ from 
lower-twist operators.
The twist-4 term contains
a twist-2 contribution,
$a_2$, and a twist-3 contribution, $d_2$, in addition to $f_2$, the 
twist-4 component\cite{SV,stein,JU,JM}:
\begin{equation}
\label{eq:mu4}
\mu_4
=M^2
\left( a_2 + 4 d_2 + 4 f_2 \right)/9 .
\end{equation}
The twist-2 matrix element $a_2$ is: 
\begin{eqnarray}
 \label{eq:a2op}
 a_2\ S^{\{ \mu} P^\nu P^{\lambda\} }
 &=& {1\over 2} \sum_q e_q^2\
     \langle P,S |
    \bar\psi_q\ \gamma^{\{ \mu} iD^\nu iD^{\lambda\} } \psi_q
     | P,S \rangle\ ,
 \end{eqnarray}
where $e_q$ is the electric charge of a quark with flavor $q$, 
S and P are the covariant spin and momentum vectors, $D^\nu$ are the 
covariant derivatives, and the parentheses
 $\{ \cdots \}$ denote symmetrization of indices.
The matrix element $a_2$ is related to the second moment of the twist-2 part 
of $g_1$:
\begin{equation}
 \label{eq:a2}
 a_2(Q^2)
 = 2 \int_0^1 dx\ x^2\ g_1(x,Q^2)\ .
 \end{equation}

Taking Eq.~(\ref{eq:a2}) as the definition of $a_2$, it is now generalized to any $Q^2$, including twist-2 and higher-twist contributions.
At low $Q^2$, the inelastic part of $a_2$, is related to 
$\gamma_{g_1}$, the generalized $g_1$ polarizability:
\begin{equation}
\label{eq:a2g}
{\overline {a_2}}(Q^2)={Q^6 \over 8\pi \alpha M^3}\gamma_{g_1}.
\end{equation}
Note that at large $Q^2$, the elastic contribution is negligible and
${\overline a_2}$ becomes $a_2$. 

The twist-3 component, $d_2$, is defined by the matrix element\cite{SV,stein,JU,JM}:
\begin{equation}
 \label{eq:d2op}
 d_2 S^{[ \mu} P^{\{\nu ]} P^{\lambda\} }
 = {1\over 8} \sum_q
     \langle P,S |
    \bar\psi_q\ g \bar F^{\{ \mu \nu} \gamma^{\lambda\} } \psi_q
     | P,S \rangle\ ,
 \end{equation}
where $g$ is the QCD coupling constant,
 $\bar F^{\mu\nu}=(1/2)e^{\mu\nu\alpha\beta}F_{\alpha\beta}$,
$F_{\alpha\beta}$ are the gluon field operators, and the parentheses
 $[ \cdots ]$ denote antisymmetrization of indices.
This matrix element is related to the second moments of the twist-3 part of 
$g_1$ and $g_2$:
\begin{eqnarray}
 \label{eq:d2}
 d_2(Q^2)
&=& \int_0^1 dx\ x^2 \Bigl (2g_1(x,Q^2)+3g_2(x,Q^2)\Bigr ) \nonumber \\
=&&\hspace {-3mm} 3 \int_0^1 dx\ x^2\Bigl (g_2(x, Q^2)-g_2^{WW}(x,Q^2)\Bigr ),
 \end{eqnarray}
where $g_2^{WW}$ is the twist-2 part of $g_2$ as derived by Wandzura and 
Wilczek\cite{WW}
\begin{equation}
\label{eq:g2ww}
g_2^{WW}(x,Q^2)=g_1(x,Q^2)+\int_x^1 dy {g_1(y,Q^2) \over y}\ .
\end{equation}
The definition of $d_2$ with Eq.~(\ref{eq:d2}) is generalized to 
all $Q^2$.
At low $Q^2$,  the inelastic part of $d_2(Q^2)$ is related to the 
polarizabilities:
\begin{equation}
\label{eq:d2in}
{\overline {d_2}}(Q^2)
={8\pi \alpha M^3 \over Q^6}(\gamma_{g_1}+{3 \over 2}
\gamma_{g_2}) 
= 
{Q^4 \over 8 M^4} \Bigl (I_1(Q^2)-I_{TT}(Q^2)+{M^2Q^2 \over \alpha}
\delta_{LT}(Q^2)\Bigr ) .
\end{equation}
At large $Q^2$, ${\overline {d_2}}$ becomes $d_2$ since the elastic
contribution becomes negligible.

The twist-4 contribution to $\mu_4$ is defined by the
matrix element
\begin{eqnarray}
\label{eq:f2op}
f_2\ M^2 S^\mu
&=& {1 \over 2} \sum_q e_q^2\
\langle N |
 g\ \bar\psi_i\ \widetilde{F}^{\mu\nu} \gamma_\nu\ \psi_i
| N \rangle\,
\end{eqnarray}
where $\widetilde{F}^{\mu\nu}$ is the dual gluon-field strength tensor.
A (generalized) definition of $f_2$ in terms of the structure functions is
\begin{equation}
 \label{eq:f2}
 f_2(Q^2) = {1 \over 2}\int_0^1 dx\ x^2 
\Bigl (7g_1(x,Q^2)+12g_2(x,Q^2)-9g_3(x,Q^2)\Bigr ), 
 \end{equation}
where $g_3$ is a parity-violating spin structure function 
which can be accessed by measurement of unpolarized lepton
scattering off a longitudinally polarized target or with neutrino scattering. 
With only $g_1$ and $g_2$ data available, $f_2$ can be extracted through 
Eqs. (\ref{eq:Gamma1}) and (\ref{eq:mu4}) if the twist-6 or higher terms are
not significant. 

The twist-3 and 4 operators describe the response of the collective
color electric and magnetic fields to the spin of the nucleon.
Expressing these matrix elements in terms of the components of
$\widetilde{F}^{\mu\nu}$ in the nucleon rest frame, one can relate
$d_2$ and $f_2$ to color electric and magnetic polarizabilities.
These are defined as\cite{SV,stein,JU,JM}
\begin{equation}
\chi_E\ 2 M^2 \vec S
= \langle N |\ \vec j_a \times \vec E_a\ | N \rangle\ , \ \ \
%
\chi_B\ 2 M^2 \vec S
= \langle N |\ j_a^0\ \vec B_a\ | N \rangle\ ,
\end{equation}
where $\vec S$ is the nucleon spin vector,
$j_a^\mu$ is the quark current,
$\vec E_a$ and $\vec B_a$ are the color electric and
magnetic fields, respectively.
In terms of $d_2$ and $f_2$ the color polarizabilities can be
expressed as
\begin{equation}
\chi_E = {2 \over 3} \left( 2 d_2\ +\ f_2 \right),  \ \ \
\chi_B = {1 \over 3} \left( 4 d_2\ -\ f_2 \right).
\label{eq:chi}
\end{equation}
\section{Summary of previous experimental situation}
Before Jefferson Lab started running polarized beams and targets,
 most of the spin structure 
measurements of the nucleon were performed at high-energy facilities like CERN (EMC\cite{EMC} and SMC\cite{SMCp,SMCd,SMCd2,SMCd0,SMCp2}), 
DESY (HERMES\cite{HERMESn,HERMESGDH,HERMESp}) and SLAC (E80\cite{E80}, E130\cite{E130}, E142\cite{E142},
E143\cite{E143}, E154\cite{E154,E154g2}, E155\cite{E155,E155g2} and E155x\cite{E155x}). 
The measured $g_1$ 
and $g_2$ data were suitable for an analysis in terms of perturbative QCD. The impetus for 
performing these experiments on both the proton and the neutron was to test the Bjorken 
sum rule, a fundamental sum rule of QCD. After twenty-five years of active 
investigation 
this goal was accomplished with a test of this sum rule to better than 10\%. 
The spin structure of the nucleon was unraveled in the same process. 
Among the 
highlights of this effort is the determination of the total spin content of the nucleon due to 
quarks, $\Delta\Sigma$ (see Eq.~(\ref{eq:mu2})). It revealed the important role
of the quark orbital angular momentum and gluon total angular momentum. 
The main results
from this inclusive double spin asymmetry measurement program have led to new directions namely the 
quest for an experimental determination of the orbital angular momentum contribution\cite{ji97} (e.g., with Deep Virtual Compton Scattering
at Jefferson Lab and other facilities) and
the gluon spin contribution (with  COMPASS\cite{COMPASS} and  RHIC-spin\cite{RHICspin} experiments). These efforts will be ongoing 
for the next decades.

\section{Description of the JLab experiments}
\label{JLabexp}

The inclusive experiments described here took place in the fall of 1998 in 
JLab Halls A\cite{HallA nim} and B\cite{HallB nim}. The
accelerator produces a CW electron beam of energy up to 6 GeV. In Hall A the 
expriment was performed using incident beam 
energies of 5.06, 4.24, 3.38, 2.58 1.72 and 0.86 GeV and a beam current up 
to 15 $\mu$A. In Hall B, beam energies 
of 2.56 and 4.28 GeV and currents 
up to 2.5 nA were used. A beam polarization of about 0.70, as measured by Halls A and B
M{\o}ller polarimeters, was 
obtained by illuminating a strained GaAs cathode with polarized laser light. 
Data taken with unpolarized targets show that the beam charge 
asymmetry and false asymmetries are under control.  
\subsection{Hall A}
\label{HallA}
A polarized high pressure ($\sim$12 atm.) gaseous $^3$He 
target\cite{jen00,kom00,e94010web} was used as an effective polarized neutron 
target in the experiment performed in Hall A. The target cell is made of two connected chambers. Rubidium atoms, 
confined by a thermal gradient to the upper spherical chamber, are 
polarized by optical pumping. The polarization is then transfered 
to $^3$He nuclei by spin-exchange collisions. The latter descend by diffusion and 
convection to the lower chamber, a 40 cm long tube of 2 cm diameter, where 
they interact with the beam.
The spins are held in a given direction by a 2.5 mT uniform field. 
The average target polarization, monitored by NMR and EPR techniques\cite{jen00,kom00,e94010web}, was 
0.35$\pm 0.014$ and its direction could be oriented longitudinal or transverse 
to the beam direction. 
The measurement of cross sections in the two orthogonal directions allowed 
a direct extraction of $g_1^{3He}$ and $g_2^{3He}$, or 
equivalently $\sigma_{TT}$  and $\sigma_{LT}$, without the need of models or
unpolarized data.

The beam energy was measured with two independent devices 
to a few $10^{-4}$ precision\cite{HallA nim}, while its current was monitored to better than 
$1\%$.
The scattered electrons were detected by two High Resolution 
Spectrometers (HRS) of $\simeq6$ msr angular acceptance and $\simeq9\%$ 
relative momentum acceptance, pointing at a nominal 15.5$^\circ$ angle. 
The high luminosity of $10^{36}$ cm$^{-2}$s$^{-1}$ allowed for 
statistically accurate data at numerous HRS momentum settings
that covered the ($Q^2$,$\nu$) plane, as is necessary to form moments at fixed $Q^2$. 
The detector package consisted of vertical drift chambers (for momentum analysis 
and vertex reconstruction), scintillation counters (data acquisition 
trigger) and \v{C}erenkov counters and lead-glass calorimeters 
(for particle identification (PID)). The $\pi^-$ were sorted from e$^-$ with an
efficiency better than 99.9\% .
Both HRS spectrometers were used to double the statistics and constrain the 
systematic uncertainties, by comparing the cross sections extracted using each HRS. 
Acceptances and optical properties were studied with a multi-foil carbon
target and 
sieve-slit collimators. Carbon elastic cross sections were measured with 2\% 
accuracy.

The kinematic coverage spans from the elastic reaction to slightly beyond
the resonance region. Data from the pion threshold to $W=2$ GeV were used to 
form neutron moments. Polarized $^3$He elastic data were used to cross-check 
target and beam
polarimetry at the 4\% level and absolute cross sections at the $5-7\%$ level. 
%
%
Electromagnetic radiative corrections were performed using the method of Mo 
and Tsai\cite{mo69} for
external radiative corrections, while internal radiative corrections were
evaluated with a modified version of the POLRAD code\cite{aku94} including 
new quasi-elastic and resonance data. The relative uncertainty on the 
radiative
corrections was typically less than 20\%. Other systematic uncertainties came from absolute 
cross-section measurements (5\%), beam and target polarizations (both 
relative 4\%). 

The spin structure functions $g_1^n$ and $g_2^n$ are extracted using polarized cross-section differences
in which contributions from unpolarized materials such as 
target windows, nitrogen or the two protons of the $^3$He nucleus cancel. 
Corrections for the two protons in $^3$He are still needed since they 
are slightly 
polarized due to the D state ($\sim$ 8\%) and S' state ($\sim$ 1.5\%) 
of the $^3$He wave function\cite{cio97}.   
Corrections for binding and Fermi motion are also applied via a PWIA-based 
model\cite{cio97}. 
The uncertainty on this correction varies from 
10\% at the lowest $Q^2$ to $< 5\%$ at larger $Q^2$. 
To form the neutron moments, the integrands (e.g. $\sigma_{TT}$ or $g_1$) are 
needed at constant $Q^2$.  Six equally spaced values of $Q^2$ were chosen 
in the range  $0.1 \le Q^2 \le 0.9\,{\rm GeV}^2$  and the integrands were 
determined from the measured points by interpolation. 
To complete the moments for the unmeasured high-energy region,
the Bianchi and Thomas parameterization\cite{tho00} was used for 
$4 < W^2 < 1000$ GeV$^2$ and a Regge-type parameterization was used 
for $W^2 > 1000$ GeV$^2$.
\subsection{Hall B}
Polarized solid $^{15}$NH$_3$ and $^{15}$ND$_3$ targets\cite{NH3} 
using dynamic nuclear polarization\cite{DNP,crabb} were used in Hall B. 
Polarizations varied from 70\% to 40\% for NH$_3$ and 25\% to 10\%  for ND$_3$.
The target material was cooled to 1.2 K and the spins were oriented by a 
longitudinal 5 T field. 

The beam was rastered to evenly spread 
depolarization effects. The beam energy was obtained from Hall A measurements. 
Its current was monitored by a Faraday cup 
downstream of the target. The luminosity reached   
$10^{33}$cm$^{-2}$s$^{-1}$. 
The CEBAF Large Acceptance Spectrometer (CLAS)
has a large angular 
($2.5 \pi$ sr) and 
momentum acceptance, which is ideal for studying the $Q^2$ evolution of 
moments. CLAS contains six 
superconducting coils that produce a toroidal magnetic field. Particles are 
characterized by three layers of drift chambers (tracking, momentum and charge 
information), a layer of scintillator paddles (time of flight), 
and \v{C}erenkov counters supplemented by electromagnetic calorimeters (PID). After 
PID, the remaining $\pi^-$ contamination was less 
than 1\% of the electron rate. 
To cover lower angles and 
momenta, the target was 
shifted 55 cm upstream of its nominal location. 
The polar 
angle ranged from 8$^o$ to 50$^o$. The azimuthal angular acceptance was 
about 80\%. The lowest momentum 
accepted was 0.35 GeV for the proton run. The deuteron data were 
taken with in-bending torus field only, which yields a minimum angle of 
14$^o$ and a minimum momentum  of 0.5 GeV. 

Since cross sections are more difficult to measure in a large-acceptance 
spectrometer and $F_2(x,Q^2)$ and 
$R(x,Q^2)$ are relatively well known\cite{F2&RSLAC,F2&RJLab}, it is easier to extract 
$g_1(x,Q^2)$ using asymmetry measurements\cite{E143}.
The physics asymmetry $A_{||}$ is related to the measured one:
$A_{raw}=A_{||} (P_b P_t DF)/C_N$.
$C_N$=0.98 is the effect of the polarized valence proton of  
$^{15}N$, $P_b$($P_t$) is the beam (target) polarization. 
$DF \simeq 0.15 (0.2)$ for 
NH$_3$ (ND$_3$) is the dilution factor from $^{15}$N, 
$^4$He and windows on the 
beam path. $DF$ was estimated using data from a carbon target. 
Parameterizations of proton and 
neutron cross sections were used to account for the difference between  
$^{15}N$ and $^{12}$C nuclei. Acceptance and detector inefficiency cancels out to first 
order in the asymmetry ratio. Acceptance cuts were chosen such that 
second-order 
corrections remain negligible. The radiative corrections to $A_{||}$ were done 
using RCSLACPOL\cite{RCSLACPOL} based on the same formalism as POLRAD. The 
cross sections are parameterized using the NMC fit of the unpolarized 
structure-function data\cite{NMC} and the polarized data from SLAC, CERN and 
HERMES\cite{Fji}. The resonance contribution is estimated by 
the MAID\cite{dre01} and the ``AO''\cite{AO} models. The radiative 
corrections account for 20\% of the total systematic uncertainty on 
$\Gamma_1$ at low $Q^2$ and 5\% at larger $Q^2$.

The same model was used to extract $A_1$ and $g_1$ from $A_{||}$. 
$F_2^p(x,Q^2)$ and $R^p(x,Q^2)$ were measured at SLAC\cite{F2&RJLab} 
and JLab\cite{F2&RJLab}. The DIS part of $g_2$ is 
calculated using  
Eq.(\ref{eq:g2ww}) while its resonance part is obtained from the 
MAID and ``AO'' models. 
The model-dependence dominates the systematic uncertainty 
on $\Gamma_1$, contributing to about 75\% at low $Q^2$ and 50\% at 
larger $Q^2$ for the proton and 50\%  in average for the deuteron. 
The model was also used to estimate the unmeasured high-energy part of 
$\Gamma_1^{p,d}$. Nuclear corrections were applied to
$\Gamma_1^{d}$ to extract $\Gamma_1^{n}$ by accounting for the deuteron 
D-state: $\Gamma_1^{n} = 2\Gamma_1^{d} / (1-1.5\omega _D)-\Gamma_1^{p}$, 
with the D-state probability $\omega_D \simeq 0.05$.
\section{Recent results from Jefferson Lab}
\subsection{Results of the generalized GDH sum for the neutron}
\label{GDHsum}
\begin{figure}[ht!]
\centerline{\psfig{figure=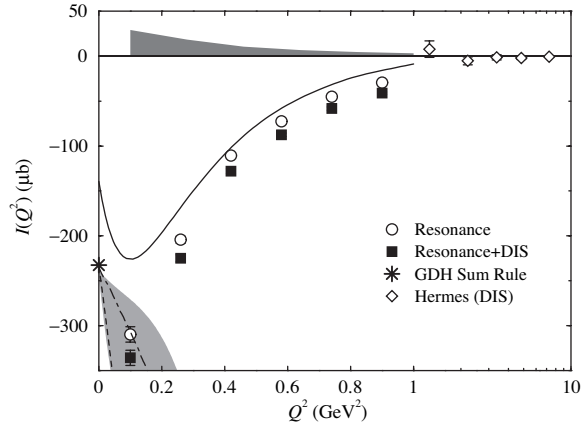,width=3.0in}}
\caption{Results for $I(Q^2)$ are compared with $\chi$PT calculations of 
ref.~\protect\cite{ji00} (dotted line) and ref.~\protect\cite{ber02} 
(dot-dashed line).  
The MAID model calculation of ref.~\protect\cite{dre01},
is shown with a solid line.  Data from HERMES\protect\cite{HERMESGDH} are also shown and a log scale is used for $Q^2 > 1$ GeV$^2$.}
\label{fig:GDH}
\end{figure}

Fig.~\ref{fig:GDH} shows the extended GDH integrals  
$I(Q^2)$ (open circles) for the neutron, 
which were extracted from JLab experiment E94-010 
(Hall A)\cite{e94010-1}, from pion threshold to $W=2.0$ GeV.
The uncertainties, when visible, represent statistics only; the systematics are shown by the grey band.   
The solid squares include an estimate of the unmeasured high-energy part.
The corresponding
uncertainty is included in the systematic uncertainty band.
These data indicate a smooth variation of $I(Q^2)$  to increasingly negative values as $Q^2$ varies from $0.9\,{\rm GeV^2}$ 
towards zero.
The data are more negative than the MAID model calculation\cite{dre01}. 
The calculation includes contributions to $I(Q^2)$ for $W \le 2\,{\rm GeV}$,  
and should thus be compared with the
open circles.  At high
$Q^2$,  the data approach the HERMES\cite{HERMESGDH} neutron results (extracted from $^3$He), which spans the range $1.28\,{\rm GeV}^2 < Q^2 < 7.25\,{\rm GeV}^2$, 
but includes only the DIS part of the GDH integral. The GDH sum rule 
prediction, $I(0)=-232.8\,\mu{\rm b}$, is indicated on Fig.~\ref{fig:GDH}, along with  
extensions to $Q^2>0$ using two $\chi$PT
calculations, one using the Heavy Baryon approximation (HB$\chi$PT) \cite{ji00} 
(dotted line) and the other Relativistic Baryon $\chi$PT (RB$\chi$PT)\cite{ber02} (dot-dashed line). Shown with a grey band is RB$\chi$PT including resonance effects\cite{ber03}, which have an associated
large uncertainty due to the resonance parameters used. The uncertainty can be seen to
encompass both the lowest $Q^2$ point and the calculation of Ref. \cite{ji00}.
At  $Q^2 = 0.3$ GeV$^2$, the prediction of Ref. \cite{ber02} is much more negative than the data. 
Further calculations as well as further measurements\footnote{JLab experiment E97-110, J.-P. Chen, A. Deur, and F. Garibaldi, spokespersons.} will help 
clarify the situation. 
\subsection{First moments of $g_1$, $g_2$ and the Bjorken sum}
\begin{figure}[ht!]
\begin{center}
\centerline{\includegraphics[scale=0.58]{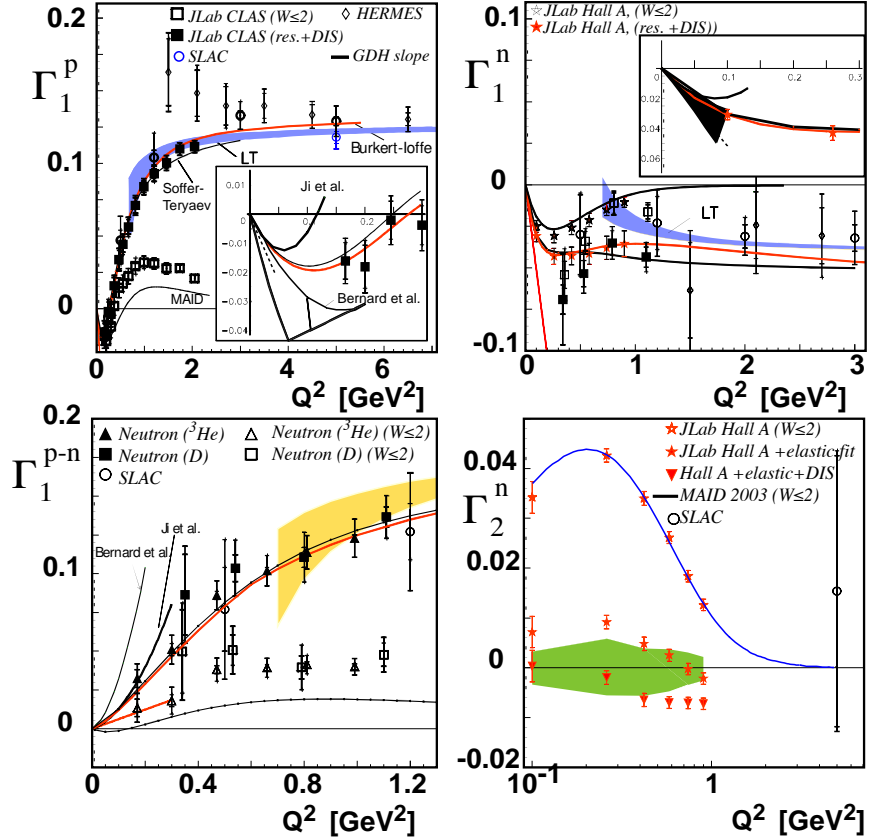}}
\end{center}
\vspace {-6mm}
\caption{ 
Results for $\overline{\Gamma}_1^p$ (top left), $\overline{\Gamma}_1^n$ (top right)
and the Bjorken sum (bottom left). The open circles (diamonds) represent the SLAC (HERMES) 
data. The open squares (stars) are the JLab CLAS (Hall A) data integrated over the resonance 
region, while the corresponding full symbols represent the full integrals. 
The slopes at $Q^2$=0 predicted by the GDH sum rule 
are given by the dotted lines. The MAID model  
predictions that includes only resonance contributions are shown by the plain 
lines while 
the dashed (dot-dashed) lines are the predictions from the Soffer-Teryaev (Burkert-Ioffe) model.
The leading twist (LT) $Q^2$-evolution of the moments is given by the grey band. The plain
lines (bands) at low $Q^2$ are the $\chi$PT 
predictions by Ji \emph{et al.} (Bernard \emph{et al.}). 
${\Gamma}_2^n$ is shown on the bottom right panel.} 
\label{fig:gamma1pn}
\vspace {-4mm}
\end{figure}

Results on $\bar \Gamma_1(Q^2)$ at low to moderate $Q^2$ are available on the proton\cite{eg1p} 
 and the neutron (using $^3$He\cite{e94010-2} and deuteron\cite{eg1d}). The integrals 
$\bar \Gamma_1^p$, $\bar \Gamma_1^n$ , $\bar \Gamma_2^n$ and $\bar \Gamma_1^{p-n}$ are 
presented in Fig.~\ref{fig:gamma1pn} along with the data (open symbols) from 
SLAC\cite{E142,E143,E154,E155,E155x} and HERMES\cite{HERMESn,HERMESp}.
The integral $\bar \Gamma_1^p$ ($\bar \Gamma_1^n$) was formed by summing 
$g_1^p$ ($g_1^n$) from pion threshold up to 
$W= 2.6$ (2.0) GeV. Nuclear corrections were applied to the deuteron and 
$^3$He data to extract $\bar \Gamma_1^n$ as described in section \ref{JLabexp}. These results are shown by the open symbols. The sums including
an estimation of the DIS contributions are shown by the solid symbols. The
inner uncertainty
indicates the statistical uncertainty while the outer one is the quadratic sum of the 
statistical and systematic uncertainties. 

Data from the proton and neutron can be combined to form the 
flavor non-singlet moment $\Gamma_1^{p}(Q^{2})-\Gamma_1^{n}(Q^{2})$ 
predicted in the Bjorken limit by the Bjorken sum rule\cite{bjo66}. It is generalized at large to moderate $Q^2$ using OPE
that yields a relatively simple leading-twist $Q^2$ evolution in which only 
non-singlet coefficients survive:
\begin{equation}
\int_{0}^{1}(g_{1}^{p}-g_{1}^{n})dx=\frac{g_{a}}{6}
[ 1-\frac{\alpha_{\rm{s}}}{\pi}-3.58\left(\frac{\alpha_{\rm{s}}}
{\pi}\right)^{2}-20.21\left(\frac{\alpha_{\rm{s}}}{\pi}\right)^{3} +...]
+O({1 \over Q^2}).
\label{eqn:bj}
\end{equation}
\noindent 
The Bjorken sum rule was verified to a level
better than 10\% at $Q^2$ = 5 GeV$^2$. 
 
Considering a flavor non-singlet object results in additional simplifications: the 
$\Delta$ resonance contribution mostly cancels out so that $\chi$PT 
predictions are expected to be more 
reliable\cite{bur01}, with possibly an extended domain of validity in $Q^2$. 
Furthermore, a flavor non-singlet moment is a convenient quantity for Lattice 
QCD calculations since disconnected diagrams, which are difficult to calculate, do not contribute. Hence, the Bjorken sum is particularly suited to provide 
benchmark tests for the three theoretical frameworks used to study the 
transition from hadronic to partonic degrees of freedom.

To form $\bar \Gamma_1^{p-n}$, the proton and neutron data were re-analyzed 
at the same $Q^2$ bins.
The unmeasured low-$x$ part of the integral was re-evaluated for 
p, d and $^3$He data sets using a prescription described in section 
\ref{HallA} with an additional constraint imposed by the 
Bjorken sum rule at $Q^2 = 5$ GeV$^2$. 

At $Q^2$=0, the GDH sum rule predicts the slopes of moments (dotted lines). The deviation from the
slopes at low $Q^2$ can be calculated with $\chi$PT. We show calculations by Ji {\it et al.}\cite{ji00} 
using the HB$\chi$PT and by Bernard {\it et al.} with\cite{ber03} and without\cite{ber02} including vector 
mesons and $\Delta$ degrees of freedom (the band shows a range of results due to the 
uncertainty in the $\gamma$N$\Delta$ form factor.) We do not show this result for the Bjorken sum 
because of the unknown correlation in the systematic uncertainties of the individual
$\overline{\Gamma}_1^p$ and $\overline{\Gamma}_1^n$. For the neutron
this calculation overlaps with a data point at $Q^2$ = 0.1 GeV$^2$ while none
of the $\chi$PT calculations agree with the proton data, of which the
lowest $Q^2$ point
is 0.17 GeV$^2$. The calculation of Ji {\it et al.} agrees with the lowest $Q^2$ data for the Bjorken sum.
At moderate and large $Q^2$  we show the MAID calculation\cite{dre01}, to be compared
with the data summed up to $W=2$ GeV. The model disagrees for 
$\overline{\Gamma}_1^p$, $\overline{\Gamma}_1^n$ and $\overline{\Gamma}_1^{p-n}$, with
the wrong sign of the slope at the photon point for $\overline{\Gamma}_1^{p-n}$ due
to its underestimate of the GDH integral for the neutron at $Q^2=0$. 
The disagreement 
indicates that
other contributions, not included in the model, are important.

The other calculations shown are by Soffer and Terayev\cite{sof02}
and by Burkert and Ioffe\cite{bur92}. Those should be compared to
the full experimental sum. The Soffer and Terayev model relies on the smooth 
$Q^2$-behavior of the $I_T=I_1+I_2$ integral to interpolate it between its
known $Q^2=0$ and large $Q^2$ value. $\overline{\Gamma}_1$ is extracted using the BC sum 
rule prediction on $\overline{\Gamma}_2$, Eq.~(\ref{eq:bc_noel}). Power corrections and
pQCD radiative corrections were recently added to the model to resolve
the initial discrepancy with the JLab data. 
The Burkert and Ioffe model relies on 
a parameterization of resonances and non-resonance background\cite{AO} supplemented by
a DIS parameterization based on vector meson dominance\cite{ans89}. Both models agree well with the data.  

 The leading-twist pQCD evolution is shown by the grey band. For 
$\overline{\Gamma}_1^{p(n)}$ we used $\Delta \Sigma$=0.15($\Delta \Sigma$=0.35), 
$a_8$=0.579 and $g_a$=1.267 to anchor the evolution (see Eqs.~(\ref{eq:mu2}) 
and (\ref{eqn:bj})).
The Bjorken sum rule sets the absolute scale for $\overline{\Gamma}_1^{p-n}$.
In all cases, the leading twist tracks the data down to
surprisingly low $Q^2$, which indicates an overall suppression of higher-twist
effects. This is quantatively discussed in section \ref{HT}. 

 The capability of transverse polarization of the Hall A $^3$He target 
allows precise measuremetns of $g_2$. 
The integral $\Gamma_2^n$  is extracted and 
plotted in the bottom-right panel of 
Fig.~\ref{fig:gamma1pn} in the measured region (open stars) and after adding the 
elastic contribution (solid stars)\cite{mer96}. 
The solid triangles
correspond to the results obtained after adding to the solid stars an estimated DIS 
contribution assuming $g_2 = g_2^{WW}$. The MAID estimate (solid line) agrees well with the resonance data. 
The positive light grey band corresponds to the total experimental systematic
errors. The negative dark band is the estimate of the systematic error 
for the low-$x$ extrapolation. The results (solid stars) are consistent with the BC 
sum rule to within 2 standard deviations over the measured $Q^2$ range. 
The SLAC E155x collaboration\cite{E155x} previously reported 
a neutron result at high $Q^2$ (solid circle), where the elastic contribution 
is negligible; the result is consistent with zero but with a rather 
large error bar. On the other hand, the proton result was reported to deviate 
from 
the BC sum rule by 3 standard deviations. The quoted uncertainty in this case is
3 times smaller than that for the neutron, but still large. 
\subsection{Spin Polarizabilities: $\gamma_0$, $\delta_{LT}$ and $d_2$ for the neutron}
The generalized spin polarizabilities provide benchmark tests
of $\chi$PT calculations at low $Q^2$ and 
of Lattice QCD calculations at moderate to high $Q^2$.
Since the generalized polarizabilities have an extra $1/\nu^2$ 
weighting compared to the first moments (GDH sum or 
$I_{LT}$), these integrals have less contribution from the large-$\nu$ 
region and converge much faster, which minimizes the uncertainty due to
the unmeasured region at large $\nu$. 

At low $Q^2$, the 
generalized polarizabilities have been evaluated with $\chi$PT 
calculations\cite{van02,ber02}.
One issue in the $\chi$PT calculations is how to properly
include the nucleon resonance contributions, especially the $\Delta$ resonance,
which usually dominates.
As was pointed out in Ref. \cite{van02,ber02}, while $\gamma_0$ is sensitive to 
resonances, $\delta_{LT}$ is insensitive to the $\Delta$ 
resonance. Measurements of the generalized spin
polarizabilities are an important step in understanding the dynamics of
QCD in the chiral perturbation region.

The first results for the neutron generalized forward
spin polarizabilities $\gamma_0(Q^2)$ and $\delta_{LT}(Q^2)$
were obtained at Jefferson Lab Hall A\cite{e94010-3}
over a $Q^2$ range from 0.1 to 0.9 GeV$^2$.
We will first focus on the low $Q^2$ region where the comparison with $\chi$PT
calculations is meaningful, and then show the complete data set.
\begin{figure}[ht!]
\begin{center}
\centerline{\includegraphics[scale=0.65, angle=0]{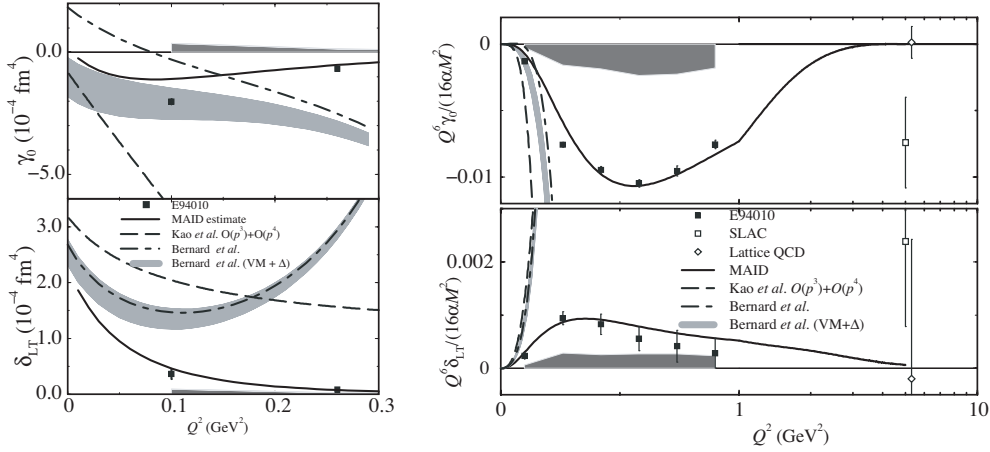}}
\end{center}
\vspace {-8mm}
\caption{Results for the neutron spin polarizabilities 
$\gamma _0$ (top panels) and $\delta _{LT}$
(bottom panels). Solid squares are the results with statistical uncertainties.
The light bands are the systematic uncertainties. 
The dashed curves are the HB$\chi $PT 
calculation\protect\cite{van02}. 
The dot-dashed curves and the dark bands are the 
RB$\chi $PT calculation\protect\cite{ber02}, 
without and with\protect\cite{ber03} 
the $\Delta $ and vector meson contributions, respectively.
Solid curves are from the MAID model\protect\cite{dre01}.
The right panels are with $Q^6$ weighting. 
The open squares are the SLAC data\protect\cite{E155x} 
and the open diamonds are the Lattice QCD calculations\protect\cite{goc01}. 
}
\label{fig:fig1}
\vspace {-3mm}
\end{figure}
The results for $\gamma_0(Q^2)$ are shown
in the top-left panel of Fig.~\ref{fig:fig1} for the two 
lowest
 $Q^2$  values of 0.10 and 0.26 GeV$^2$. 
The statistical uncertainties are smaller than the
size of the symbols. 
The data are compared with 
a next-to-leading order, $O(p^4)$, HB$\chi$PT 
calculation\cite{van02}, a next-to-leading-order RB$\chi$PT
calculation\cite{ber02}, and the same calculation explicitly including 
both the $\Delta$ resonance and vector meson contributions\cite{ber03}.
Predictions from the MAID model\cite{dre01} are also shown.
At the lowest $Q^2$ point,
the RB$\chi$PT
calculation including the resonance contributions
is in good agreement with the experimental result.
For the HB$\chi$PT calculation without explicit resonance contributions, 
discrepancies are large even at $Q^2 = 0.1$ GeV$^2$. 
This might indicate the significance of the resonance contributions or a
problem with the heavy baryon approximation at this $Q^2$.
The higher $Q^2$ data point is in good agreement with the MAID
prediction,
but the lowest data point at $Q^2 = 0.1 $ GeV$^2$ is significantly lower, 
consistent with what was observed for the generalized GDH
integral result (Section~\ref{GDHsum}).
Since $\delta_{LT}$ is 
insensitive to the dominating
$\Delta$ resonance contribution, it was believed that $\delta_{LT}$ should be
more suitable than $\gamma_0$ to serve as a testing ground for the chiral 
dynamics of QCD\cite{ber02,van02}.
The bottom-left panel of Fig.~\ref{fig:fig1} shows $\delta_{LT}$ 
compared to
$\chi$PT calculations and the MAID predictions. It is surprising to see
that 
the data are in significant disagreement with the $\chi$PT calculations 
even at the lowest $Q^2$, 0.1 GeV$^2$. 
This presents a significant challenge to the present theoretical
understanding. 
The MAID predictions are in good agreement with the results.

The right panels of Fig.~\ref{fig:fig1} show $\gamma_0$ and
$\delta_{LT}$ multiplied by $Q^6$ along with 
the MAID and $\chi$PT calculations. Also shown are the world
data\cite{E155x} and a quenched Lattice QCD 
calculation\cite{goc01}, both at $Q^2= 5$~GeV$^2$.
It is expected that
at large $Q^2$,
the $Q^6$-weighted spin polarizabilities become independent of $Q^2$
(scaling)\cite{dre03,dre04}, and the DIS
Wandzura-Wilczek 
relation\cite{WW} leads to a relation 
between $\gamma_0$ and $\delta_{LT}$:

\begin{equation}
\delta_{LT}(Q^2) \rightarrow 
{\frac{1}{3}}\gamma_0(Q^2) \ \ \ {\rm as} \ \ Q^2 \rightarrow \infty. 
\end{equation}

\noindent 
The results show that 
scaling is not observed for $Q^2 < 1$ GeV$^2$.
For the higher moments scaling is expected to 
start at a higher $Q^2$ than for the first moments, for which
scaling was observed to start around $Q^2$ of 1 GeV$^2$, 
where higher-twist effects become insignificant.
Again, both results are in good agreement with the MAID model.

\begin{figure}[ht!]
\begin{center}
\centerline{\includegraphics[scale=0.62,angle=0]{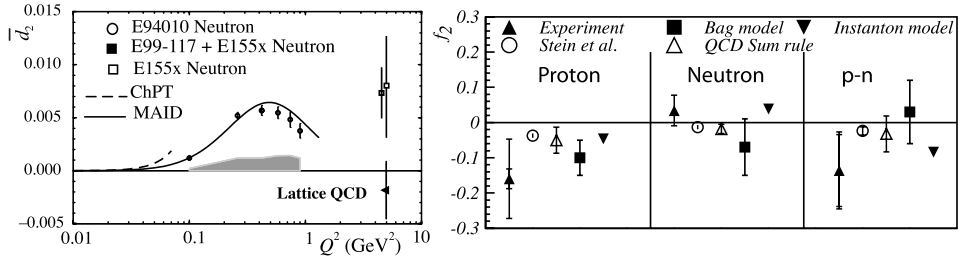}}
\end{center}
\vspace{-5mm}
\caption{The left panel shows the JLab results of $\overline d_2$ for the neutron 
along with the world data at high $Q^2$, Lattice QCD, MAID model and HB$\chi$PT
calculations. The right plot compares 
$f_2$ with calculations of Stein {\it et al.}\protect\cite{stein}, MIT
bag model\protect\cite{JM}, QCD sum rules\protect\cite{sumrule} and an instanton 
model\protect\cite{instanton}.}
\label{fig:d2}
\vspace{-5mm}
\end{figure}

Another combination of the second moments, $d_2(Q^2)$, provides an 
efficient way to study the high $Q^2$ behavior of the nucleon spin structure,
since it is a matrix element, related to the color polarizabilities and can 
be calculated from Lattice QCD. It also provides a means to study the 
transition from high to low $Q^2$. 
In Fig.~\ref{fig:d2}, $\bar d_2(Q^2)$  
is shown. The experimental results are the solid circles. 
The grey band represents the systematic uncertainty. The world neutron results from\cite{E155x}(open square) and from JLab E99-117\cite{Zheng} 
(solid square) are also shown. The solid line is the
MAID calculation containing only the resonance contribution. 
At low $Q^2$ the HB$\chi$PT calculation\cite{van02} (dashed line) is shown.
The RB$\chi$PT\cite{ber02} 
with or without the vector mesons  
and the $\Delta$ contributions\cite{ber03} 
are very close to the HB$\chi$PT curve at this scale, and are not shown on the
figure for clarity.
The Lattice QCD prediction\cite{goc01} at $Q^2$ = 5~GeV$^2$ is negative but close to zero. We note that all models (not shown at
this scale) predict a negative or zero value at large $Q^2$. At moderate $Q^2$,
our data show that $\bar d_2^n$ is positive and decreases with $Q^2$.

\subsection{Higher Twist extractions and $f_2$}
\label{HT}
The higher-twist contribution to $\Gamma_1$ can be obtained
by a fit with an OPE series, Eq.~(\ref{eq:Gamma1}), truncated to an order
appropriate for the precision of the data.
The goal is to determine the twist-4 matrix element $f_2$.
Once $\mu_4$ is obtained, $f_2$ is extracted 
by subtracting the leading-twist contributions of $a_2$ and $d_2$ 
following Eq.~(\ref{eq:mu4}). To have an idea how the higher-twist terms
(twist-8 and above) affect the twist-4 term extraction,
it is necessary to study the convergence of the expansion 
and to choose the $Q^2$ range in a region where $\mu_8$ term is not
significant. This study is made possible only with the 
availability of the new low-$Q^2$ data from JLab.

Higher-twist analyses 
have been performed on the proton\cite{HTP2}, the neutron\cite{HTN} and the Bjorken 
sum\cite{BJ}. An earlier proton analysis is available\cite{HTP} but will not 
be presented here since it uses a different procedure.
$\Gamma_1$ at moderate $Q^2$ is obtained as described in section~\ref{JLabexp}.
For consistency, the unmeasured low-$x$ parts of the JLab $\Gamma_1^p$ and 
of the world data on $\Gamma_1$ were re-evaluated using the same prescription 
previously used for $\Gamma_1^n$ and $\Gamma_1^{p-n}$. 
The elastic contribution, negligible above 
$Q^2$ of 2 GeV$^2$ but significant (especially for the proton) at lower values 
of $Q^2$, was added using the parametrization of Ref.~\cite{mer96}.
The leading-twist term $\mu_2$ was determined by fitting the 
data at $Q^2 \ge 5$ GeV$^2$ assuming that higher twists
in this $Q^2$ region are negligible. A value of
$g_A=1.270 \pm 0.045$ was obtained for the Bjorken sum.
Using the proton (neutron) data alone, and with input of 
$g_A$ from the neutron beta decay and $a_8$ from the hyperon 
decay (assumed SU(3) flavor symmetry), 
we obtained $\Delta\Sigma=0.15 \pm 0.07$ for the proton and
$\Delta\Sigma=0.35\pm 0.08$ for the neutron.  We note that there 
is a significant difference between $\Delta\Sigma$ determined from the proton 
and from the neutron data.
This is the main reason why the extracted $\mu_4$
and $f_2$ from the Bjorken sum is different compared to the difference of those 
extracted individually from the proton and neutron, since the Bjorken sum does not 
need the assumption of SU(3) flavor symmetry and $\Delta \Sigma$ was cancelled.

The fit results using an expansion up to ($1/Q^6$) in determining $\mu_4$ are 
summarized in Table~\ref{table1}. 
In order to extract $f_2$, shown in Table~\ref{table2},
the target-mass corrections $a_2$ were evaluated using the Blumlein-Boettcher world data parametrization\cite{BB} for the proton and a world fit to the neutron data, which includes the recent
high precision neutron results at large $x$\cite{Zheng}. The $d_2$ values 
used are from SLAC E155x\cite{E155x} (proton) and JLab E99-117\cite{Zheng}
(neutron). 
\begin{table}[ht]
\tbl{Results of $\mu_4$, $\mu_6$ and 
$\mu_8$ at $Q^2$ = 1 GeV$^2$ for proton, neutron and $p-n$. The uncertainties are
first statistical then systematic.} 
{\begin{tabular}{|c|c|c|c|c|}
\hline 
Target& $Q^{2}$ (GeV$^{2}$)&
$\mu_4/M^2$&
$\mu_6/M^4$&
$\mu_8/M^6$\tabularnewline
\hline
\hline 
proton & 0.6-11.0&
-0.065$\pm 0.012 \pm 0.048$&
0.143$\pm 0.021 \pm 0.056$&
-0.026$\pm 0.008 \pm 0.016$\tabularnewline
neutron&0.5-11.0&
0.019$\pm 0.002 \pm 0.024$&
-0.019$\pm 0.002 \pm 0.017$&
0.00$\pm 0.00 \pm 0.03$\tabularnewline
$p-n$&0.5-11.0&
-0.060$\pm 0.045 \pm 0.018$&
0.086$\pm 0.077 \pm 0.032$&
0.011$\pm 0.031 \pm 0.019$\tabularnewline
\hline
\end{tabular}}
\label{table1}
\end{table}
\begin{table}[ht]
\tbl{Results of $f_2$, $\chi_E$ and 
$\chi_B$ at $Q^2$ = 1 GeV$^2$ for proton, neutron and $p-n$.
The uncertainties are
first statistical then systematic.} 
{\begin{tabular}{|c|c|c|c|}
\hline 
Target & $f_2$ & $\chi_E$ & $\chi_B$ \tabularnewline
\hline
\hline 
$p$ & -0.160 $\pm 0.028 \pm 0.109$ & -0.082 $\pm 0.016 \pm 0.071$ &
0.056 $\pm 0.008 \pm 0.036$\tabularnewline
$n$ & 0.034 $\pm 0.005 \pm 0.043$ & 0.031 $\pm 0.005 \pm$ 0.028 & 
-0.003 $\pm$ 0.004 $\pm$ 0.014 \tabularnewline
$p-n$ & $-0.136 \pm 0.102 \pm 0.039$ & $-0.100 \pm 0.068 \pm 0.028 $
& $0.036 \pm 0.034 \pm 0.017 $ \tabularnewline
\hline
\end{tabular}}
\label{table2}
\end{table}

The $\Gamma_1$ were fit, varying the minimum $Q^2$ threshold 
to study the convergence of the OPE series. 
The extracted quantities have large uncertainties (dominated by the 
systematic uncertainty) but are stable with respect to the minimal $Q^2$ 
threshold when it was below 1 GeV$^2$. 
The results do not vary significantly when the $
\mu_8$ term is added, which justifies \emph{a posteriori} the use of the 
truncated OPE series in the chosen $Q^2$ range. In the proton case, the 
elastic contribution makes a significant contribution to the $\mu_6$ term at 
low $Q^2$ but this does not invalidate \emph{a priori} the validity of the
series since the elastic contributes mainly to $\mu_6$ and $\mu_8$ remains
small compared to $\mu_4$. We notice the alternation of signs between the 
coefficients. This leads to a partial suppression of the higher-twist 
effects and may be a reason for quark-hadron duality in
the spin sector\cite{duality}. We also note that the sign 
alternation is opposite for the proton and neutron. The results are 
compared to theoretical calculations in Fig. \ref{fig:d2}.
Following Eq.~(\ref{eq:chi}), the electric and magnetic color polarizabilities were determined. 
Overall, the values, given 
in Table~\ref{table2}, are small, and we observe a sign change 
in the electric color polarizability between the proton and the neutron. 
We also expect a sign change in the color magnetic polarizability. 
However, with the large uncertainty and the 
small negative value of the neutron $\chi_B$, it is difficult to confirm 
this expectation.

\section{Conclusion}

A large body of nucleon spin-dependent cross-section and 
asymmetry data have been collected at low to moderate $Q^2$ in the resonance region. 
These data have been used to evaluate the $Q^2$ evolution of moments of 
the nucleon spin structure functions $g_1$ and $g_2$, including the 
GDH integral, the Bjorken sum, the BC sum and the spin polarizabilities, 
and to extract higher-twist contributions. The latter 
provided access to the color polarizabilities in the nucleon. 
  
At $Q^2$ close to zero, available next-to-leading order $\chi$PT calculations
were tested against the data and found to be in reasonable agreement for $Q^2$ =
0.1 GeV$^2$ for the GDH integral $I(Q^2)$, $\Gamma_1(Q^2)$ and the  forward spin
polarizability $ \gamma_0(Q^2)$. Above $Q^2$ =0.1 GeV$^2$  a significant 
difference between the calculation and the data is observed, pointing to the
limit of applicability of $\chi$PT as $Q^2$ becomes larger. Although it was
expected that the $\chi$PT calculation of $\delta_{LT}$ would offer a faster
convergence because of the absence of the $\Delta$ contribution, the
experimental data show otherwise. None of the available $\chi$PT
calculations can  reproduce $\delta_{LT}$ at $Q^2$ = 0.1 GeV$^2$.  This discrepancy presents a significant challenge
to our theoretical understanding at its present level of approximations
and might indicate that higher order calculations are necessary. 

Overall, the trend of the data  is well described by phonemenological 
models. The dramatic $Q^2$ evolution of $I_{GDH}$ from high to  low $Q^2$ was
observed as predicted by these models for both the proton and the neutron. This 
behavior is mainly determined by the relative strength and sign of the $\Delta$ 
resonance compared to that of higher energy resonances and deep inelastic processes.
This also shows that the current level of phenomenological understanding of the
resonance spin structure using these moments as observables is reasonable. 

The neutron BC sum rule is observed to be verified within
uncertainties in the
low $Q^2$ range  due to a cancellation between the resonance and the
elastic contributions. The BC sum rule is expected to be valid at all $Q^2$.
This test validates the assumptions going ino the BC sum rule,
which provides confidence in sum rules with similar assumptions.

In the $Q^2$ region above 0.5 GeV$^2$ the 
first moment of $g_1$ for the proton, neutron and the proton-neutron 
difference were re-evaluated using the world data and the same extrapolation 
method of the unmeasured regions
for consistency. Then, in the framework of the OPE, the higher twist $f_2$ and $d_2$ matrix 
elements were extracted. The low $Q^2$ data allowed us to gauge the convergence of the 
expansion used in this analysis. The extracted higher-twist (twist-4 and above) effects are not significant for $Q^2$ above 1 GeV$^2$. This fact may be related to the observation that quark-hadron duality works reasonably well 
at Q$^2$ above 1 GeV$^2$.

Finally, the proton and neutron electric and magnetic 
color polarizabilities were determined by combining the twist-4 matrix element $f_2$
and the twist-3 matrix element $d_2$ from the world data. Our findings show 
 a small and slightly positive value of $\chi_E$ and a value of $\chi_B$ close to zero 
for the neutron, while the proton has slightly larger values for both 
$\chi_E$ and $\chi_B$ but with opposite signs.

Overall, 
the new JLab data have provided valuable information on the transition between the non-perturbative 
to the perturbative regime of QCD. They form a precise data set for twist
expansion analysis and a check of $\chi$PT calculations.

Future precision measurements\footnote{JLab proposal PR03-107, S. Choi, X. Jiang, Z.-E., Meziani, spokespersons.}\cite{12gev} of the
$g_1$ and $g_2$ structure functions at $Q^2 \approx 1-4$~GeV$^2$
will reduce the uncertainty in the extracted higher-twist coefficients
and provide a benchmark test of Lattice QCD.

\section*{Acknowledgments}
Thanks to Kees de Jager, Ron Gilman and Vince Sulkosky for careful
proof-reading. This work was supported by the U.S. Department of Energy (DOE).
The Southeastern Universities
Research Association operates the Thomas Jefferson National Accelerator
Facility for the DOE under contract DE-AC05-84ER40150, Modification No. 175.



\begin{thebibliography}{0}
\bibitem{Fji} see, {\it e.g.}, B. W. Filippone and X. Ji, Adv. Nucl. Phys. {\bf 26}, 1 (2001).
\bibitem{bjo66} J. D. Bjorken, Phys. Rev. {\bf 148}, 1467 (1966).
\bibitem{e94010-1} M. Amarian {\em et al.}, Phys. Rev. Lett. {\bf 89}, 242301 (2002).
\bibitem{e94010-2} M. Amarian {\em et al.}, Phys. Rev. Lett.
{\bf 92}, 022301 (2004).
 \bibitem{e94010-3} M. Amarian {\em et al.}, Phys. Rev. Lett. {\bf 93}, 152301 (2004).
\bibitem{eg1p} R. Fatemi {\em et al.}, Phys. Rev. Lett. {\bf 91}, 222002 (2003).
\bibitem{eg1d} J. Yun {\em et al.}, Phys. Rev. {\bf C 67}, 055204 (2003). 
\bibitem{HTN} Z.-E. Meziani {\em et al.}, Phys. Lett. {\bf B 613}, 148 (2005).
\bibitem{HTP} M. Osipenko {\em et al.}, Phys. Lett. {\bf B 609}, 259 (2005).
\bibitem{HTP2} A. Deur, nucl-exp/0508022.
\bibitem{BJ} A. Deur {\em et al.}, Phys. Rev. Lett. {\bf 93}, 212001 (2004).
\bibitem{GDH} S. B. Gerasimov, Sov. J. Nucl. Phys. {\bf 2}, 598 (1965)
\bibitem{GDH2} S. D. Drell and A. C. Hearn, Phys. Rev. Lett. {\bf 16}, 908 (1966).
\bibitem{ahr01} J. Ahrens {\it et al.} Phys. Rev. Lett. {\bf 87}, 022003 (2001).\bibitem{dutz03}
 H. Dutz {\it et al.}, Phys. Rev. Lett. {\bf 91}, 192001 (2003).
\bibitem{dutz04}
 H. Dutz {\it et al.}, Phys. Rev. Lett. {\bf 93}, 032003 (2004).
\bibitem{dre01} D. Drechsel, S.S. Kamalov, and L. Tiator, Phys. Rev. {\bf D 63}, 114010 (2001). 
\bibitem{ji01} X. Ji and J. Osborne, J. of Phys. G
{\bf 27}, 127 (2001).
\bibitem{ans89} M. Anselmino, B. L. Ioffe, and E. Leader, Sov. J. Nucl. Phys. {\bf 49}, 136 (1989).
\bibitem{dre03} D. Drechsel, B. Pasquini and M. Vanderhaeghen, Phys. Rep. 
{\bf 378}, 99 (2003).
\bibitem{dre04} D. Drechsel and L. Tiator, Ann. Rev. Nucl. Part. Sci.
{\bf 54}, 69 (2004).
\bibitem{BD} see, {\it e.g.}, J. D. Bjorken and S. D. Drell, ``Relativistic Quantum Fields'',
McGraw Hill, New York (1965). 
\bibitem{low} F. E. Low, Phys. Rev. {\bf 96}, 1428 (1954).
\bibitem{ope} K. Wilson, Phys. Rev. {\bf 179}, 1499 (1969).
\bibitem{BC} H. Burkhardt and W. N. Cottingham, Ann. Phys. (N.Y.) {\bf 56}, 453 (1970).
\bibitem{EJ} J. Ellis and R. L. Jaffe, Phys. Rev. {\bf D 9}, 1444 (1974); 
{\it ibid.}, {\bf D 10}, 1669 (1974).
\bibitem{SV}E.~V.~Shuryak and A.~I.~Vainshtein,
{\it Nucl. Phys.} {\bf B 201}, 141 (1982). 
\bibitem{stein} E.~Stein, P.~Gornicki, L.~Mankiewicz and A.~Sch\"afer,
{\it Phys. Lett.} {\bf B 353}, 107 (1995).
\bibitem{JU} X. Ji and P. Unrau, Phys. Lett. {\bf B 333}, 228 (1994). 
\bibitem{JM} X. Ji and W. Melnitchouk, Phys. Rev. {\bf D 56},1 (1997).
\bibitem{WW} S. Wandzura and F. Wilczek, Phys. Lett.  {\bf B 72}, 195 (1977).
\bibitem{EMC} J. Ashman {\em et al.}, Phys. Lett. {\bf B 206}, 364 (1988). 
\bibitem{SMCp} SMC collaboration: D. Adams \emph{et al.}, Phys. Lett. {\bf B 329}, 399 (1994).
\bibitem{SMCd} SMC collaboration: D. Adams \emph{et al.}, Phys. Lett. {\bf B 357}, 248 (1995).
\bibitem{SMCd2} SMC collaboration: D. Adams \emph{et al.}, Phys. Lett.{\bf B 396}, 338 (1997).
\bibitem{SMCd0} SMC collaboration: D. Adeva \emph{et al.}, Phys. Lett. {\bf B 302}, 533 (1993).
\bibitem{SMCp2} SMC collaboration: D. Adeva \emph{et al.}, Phys. Lett.{\bf B 412}, 414 (1997).
\bibitem{HERMESn} K. Ackerstaff {\em et al.}, Phys. Lett. {\bf B 404}, 383 (1997).
\bibitem{HERMESGDH} K. Ackerstaff {\em et al.}, Phys. Lett. {\bf B 444}, 531 (1998).
\bibitem{HERMESp} A. Airapetian, {\em et al.}, Phys. Lett. {\bf B 442}, 484 (1998).
\bibitem{E80} M. J. Alguard {\em et al.}, Phys. Rev. Lett. {\bf 37}, 1261 (1976).
\bibitem{E130} M. J. Alguard {\em et al.}, Phys. Rev. Lett. {\bf 41}, 70 (1976).
\bibitem{E142} P. L. Anthony {\em et al.}, Phys. Rev. {\bf D 54}, 6620 (1996).
\bibitem{E143} K. Abe {\em et al.},
Phys. Rev. {\bf D 58},112003 (1998). 
\bibitem{E154} K. Abe {\em et al.},
Phys. Rev. Lett. {\bf 79}, 26 (1997).
\bibitem{E154g2} K. Abe {\em et al.}, Phys. Lett. {\bf B 404}, 377 (1997).
\bibitem{E155} P. L. Anthony, {\em et al.}, Phys. Lett. {\bf B 493}, 19 (2000).
\bibitem{E155g2} P. L. Anthony, {\em et al.}, Phys. Lett. {\bf B 458}, 529 (1999).
\bibitem{E155x} P. L. Anthony {\em et al.}, Phys. Lett. {\bf B 553}, 18 (2003).
\bibitem{COMPASS} C. Bernet, Proceedings of DIS2005 (2005), Madison, Wisconsin,
AIP Conf. Proc. (2005).
\bibitem{RHICspin} A. Deshpande, Proceedings of DIS2005 (2005), Madison, Wisconsin, AIP Conf. Proc. (2005).
\bibitem{ji97} X. Ji, Phys. Rev. Lett. {\bf 78}, 610 (1997).
\bibitem{HallA nim}
Hall A collaboration: J. Alcorn \emph{et al.}, 
Nucl. Inst. Meth. \textbf{A522}, 294 (2004).
\bibitem{HallB nim}
CLAS collaboration: B. A. Mecking \emph{et al.},
Nucl. Inst. Meth. \textbf{A503}, 513 (2003).
\bibitem{jen00} J.S. Jensen, Ph.D. Thesis, California Institute of Technology, 2000 (unpublished)
\bibitem{kom00} I. Kominis, Ph.D. Thesis,
Princeton University, 2000 (unpublished).
\bibitem{e94010web} Details of JLab E94-010 and relevant theses can be found at www.jlab.org/e94010/.
\bibitem{mo69} L.W. Mo and Y.S. Tsai, Rev. Mod. Phys. {\bf 41}, 205 (1969).
\bibitem{aku94} I.V Akushevich and N.M. Shumeiko, J. Phys.  {\bf G 20}, 513 (1994).
\bibitem{cio97} C. Ciofi degli Atti and S. Scopetta, Phys. Lett. {\bf B 404}, 223 (1997). 
\bibitem{DNP} 
A. Abragam and M. Goldman, Rep. Prog. Phys. \textbf{41} 396 (1978).
\bibitem{crabb}
D.G. Crabb and W. Meyer, Annu. Rev. Nucl. Part. Sci. \textbf{47}, 67 (1997).
\bibitem{NH3} 
C. D. Keith \emph{et al}., Nucl. Inst. Meth. A \textbf{501}, 327 (2003).
\bibitem{F2&RSLAC} L.W. Whitlow {\it et al.}, Phys. Lett. {\bf B 250}, 193 (1990);
\bibitem{F2&RJLab}
Y. Liang et al., nucl-ex/0410027.
\bibitem{RCSLACPOL} 
K. Abe \emph{et al}., Phys. Rev. \textbf{D 58} 112003 (1998).
\bibitem{NMC} M. Arneodo \emph{et al}.,  Phys. Lett. {\bf B 364}, 107 (1995).
\bibitem{AO} V. Burkert and Z. Li, Phys. Rev \textbf{D 47}, 46 (1993).
\bibitem{Sofferbound} J. Soffer, Phys. Rev. Lett. \textbf{74}, 1292 (1995).
\bibitem{tho00} N. Bianchi and  E. Thomas, Nucl. Phys. {\bf B 82} (Proc. Suppl.), 256 (2000).
\bibitem{ji00} X. Ji, C. Kao, and J. Osborne, Phys. Lett. B {\bf 472}, 1 (2000).\bibitem{ber02} V. Bernard, T. Hemmert and Ulf-G. Meissner, Phys. Lett. 
{\bf B 545}, 105 (2002)
\bibitem{ber03} V. Bernard, T. Hemmert and Ulf-G. Meissner,
Phys. Rev. {\bf D 67}, 076008 (2003).
\bibitem{sof02}J. Soffer and O. V. Teryaev, Phys. Rev. {\bf D 70}, 116004 (2004).
\bibitem{bur92}V. D. Burkert and B. L. Ioffe, Phys. Lett. {\bf B 296}, 223 (1992).
\bibitem{12gev}
The Science Driving the 12 GeV Upgrade of CEBAF, \\
http://www.jlab.org/div\_dept/physics\_division/GeV.html.
\bibitem{mer96} P. Mergell, Ulf-G. Meissner and D. Drechsel, Nucl. Phys.  {\bf A 596}, 367 (1996).
\bibitem{mel03} S. A. Kulagin and W. Melnitchouk (to be published), private communication. 
\bibitem{van02} C. W. Kao, T. Spitzenberg and M. Vanderhaeghen, Phys. Rev. {\bf D  67}, 016001 (2003).
\bibitem{goc01} M. Gockeler {\em et al.}, Phys. Rev. {\bf D 63}, 074506, (2001).
\bibitem{bur01}
V. D. Burkert,
Phys. Rev. \textbf{D 63}, 097904 (2001).
\bibitem{BB} J. Blumlein and H. Boettcher, Nucl. Phys. \textbf{636}, 225 (2002).
\bibitem{Zheng} X. Zheng {\em et al.} Phys. Rev. {\bf C 70}, 065207 (2004). 
\bibitem{duality} W. Melnitchouk, R. Ent and C. Keppel, Phys. Rept. {\bf 406},
127 (2005).
\bibitem{sumrule} I.I. Balitsky, V. M. Braun and A.V.
Kolesnichenko, Phys. Lett. \textbf{B 242}, 245 (1990).
\bibitem{instanton} N-Y. Lee, K. Goeke and C. Weiss, Phys. Rev.
\textbf{D 65}, 054008 (2002).
\end{thebibliography}
\end{document}